\documentclass[sigplan,screen]{acmart}

\AtBeginDocument{%
  \providecommand\BibTeX{{%
    \normalfont B\kern-0.5em{\scshape i\kern-0.25em b}\kern-0.8em\TeX}}}

\acmSubmissionID{503}
\renewcommand\footnotetextcopyrightpermission[1]{}

\usepackage[binary-units=true]{siunitx}
\usepackage{glossaries}
\usepackage{graphicx}
\usepackage{multicol}
\usepackage{tabularx}
\usepackage{multirow}
\usepackage{caption}
\usepackage{subcaption}
\usepackage{cleveref}
\usepackage[shortlabels]{enumitem}

\newcommand{\encircled}[2][0.8mm]{%
    \raisebox{.5pt}{%
        \textcircled{%
            \raisebox{0.35pt}{%
                \kern #1
                \scalebox{0.70}{#2}
            }%
        }%
    }%
}

\definecolor{ourgreen}{rgb}{0.00,0.49,0.19}
\definecolor{ourred}{rgb}{0.77,0.03,0.11}
\definecolor{ourorange}{rgb}{0.89,0.45,0.13}
\definecolor{ourgrey}{rgb}{0.60,0.60,0.60}
\def\yes{\textcolor{ourgreen}{\large\checkmark}}
\def\maybe{\textcolor{ourorange}{\Large$\circ$}} 
\def\no{\textcolor{ourred}{\Large\texttimes}}
\def\unknown{\textcolor{ourgrey}{\encircled[1mm]{?}}}

\DeclareSIUnit[per-mode=symbol]\bps{\bit\per\second}
\DeclareSIUnit[per-mode=symbol]\kbps{\kilo\bps}
\DeclareSIUnit[per-mode=symbol]\Mbps{\mega\bps}
\DeclareSIUnit[per-mode=symbol]\Gbps{\giga\bps}
\DeclareSIUnit[per-mode=symbol]\nanosec{\nano\second}
\DeclareSIUnit\microsec{\SIUnitSymbolMicro s}
\DeclareSIUnit\byte{B}
\DeclareSIUnit\bit{bit}
\DeclareSIUnit\terabyte{TB}

\newcommand{\engine}{\gls{engine}}
\newcommand{\phaseinstall}{\textbf{1. Install}}
\newcommand{\phasesetup}{\textbf{2. Setup}}
\newcommand{\phasescenario}{\textbf{3. Scenario}}
\newcommand{\phaseprocess}{\textbf{4. Process}}

\newcommand{\apnode}{$\rho$-node}
\newcommand{\annode}{$\eta$-node}

\newcommand{\reqi}{\textit{\textbf{R1: Repeatability}}}
\newcommand{\reqii}{\textit{\textbf{R2: Reproducibility}}}
\newcommand{\reqiii}{\textit{\textbf{R3: Replicability}}}
\newcommand{\reqiv}{\textit{\textbf{R4: Openness}}}

\newcommand{\reqvi}{\textit{\textbf{R6: Autonomy}}}
\newcommand{\reqvii}{\textit{\textbf{R7: Malfunction Scenarios}}}

\newcommand{\reqix}{\textit{\textbf{R9: Granular Control}}}

\newcommand{\reqxii}{\textit{\textbf{R12: Scalability}}}

\newcommand{\reqxiv}{\textit{\textbf{R14: Diversity}}}
\newcommand{\reqxv}{\textit{\textbf{R15: Standardized}}}

\newcommand{\sreqi}{\textit{\textbf{R1: Repeat}}}
\newcommand{\sreqii}{\textit{\textbf{R2: Reproduce}}}
\newcommand{\sreqiii}{\textit{\textbf{R3: Replicate}}}
\newcommand{\sreqiv}{\textit{\textbf{R4: Openness}}}

\newcommand{\sreqvi}{\textit{\textbf{R6: Autonomy}}}
\newcommand{\sreqvii}{\textit{\textbf{R7: MalScenario}}}

\newcommand{\sreqix}{\textit{\textbf{R9: Gran.Con.}}}

\newcommand{\sreqxii}{\textit{\textbf{R12: Scalable}}}

\newcommand{\sreqxiv}{\textit{\textbf{R14: Diversity}}}
\newcommand{\sreqxv}{\textit{\textbf{R15: Standard}}}

\newcommand{\nreqi}{\textit{\textbf{R1}}}
\newcommand{\nreqii}{\textit{\textbf{R2}}}
\newcommand{\nreqiii}{\textit{\textbf{R3}}}
\newcommand{\nreqiv}{\textit{\textbf{R4}}}

\newcommand{\nreqvi}{\textit{\textbf{R6}}}
\newcommand{\nreqvii}{\textit{\textbf{R7}}}

\newcommand{\nreqix}{\textit{\textbf{R9}}}

\newcommand{\nreqxii}{\textit{\textbf{R12}}}

\newcommand{\nreqxiv}{\textit{\textbf{R14}}}
\newcommand{\nreqxv}{\textit{\textbf{R15}}}

\newcommand{\sysname}{\gls{methoda}}

\begin{document}

\title{Multilayer Environment and Toolchain for Holistic NetwOrk Design and Analysis}

\author{Filip Rezabek, Kilian Glas, Richard von Seck, Achraf Aroua, Tizian Leonhardt, and Georg Carle} 
  \affiliation{%
  \institution{TUM School of Computation, Information, and Technology, Technical University of Munich}
  \country{Germany} 
    }

\renewcommand{\shortauthors}{Rezabek, et al.}

\begin{abstract}
The recent developments and research in distributed ledger technologies and blockchain have contributed to the increasing adoption of distributed systems. 
To collect relevant insights into systems' behavior, we observe many evaluation frameworks focusing mainly on the system under test throughput. 
However, these frameworks often need more comprehensiveness and generality, particularly in adopting a distributed applications' cross-layer approach.
This work analyses in detail the requirements for distributed systems assessment. We summarize these findings into a structured methodology and experimentation framework called METHODA.
Our approach emphasizes setting up and assessing a broader spectrum of distributed systems and addresses a notable research gap.
We showcase the effectiveness of the framework by evaluating four distinct systems and their interaction, leveraging a diverse set of eight carefully selected metrics and 12 essential parameters. Through experimentation and analysis we demonstrate the framework's capabilities to provide valuable insights across various use cases. For instance, we identify that a combination of Trusted Execution Environments with threshold signature scheme FROST introduces minimal overhead on the performance with average latency around \SI{40}{\ms}. We showcase an emulation of realistic systems behavior, e.g., Maximal Extractable Value is possible and could be used to further model such dynamics.  
The METHODA framework enables a deeper understanding of distributed systems and is a powerful tool for researchers and practitioners navigating the complex landscape of modern computing infrastructures.

\end{abstract}

\keywords{Evaluation, Distributed Systems, Reproducability, Testbed}
\maketitle
\newacronym{nic}{NIC}{Network interface card}
\newacronym{iiot}{IIoT}{Industrial Internet of Things}
\newacronym{iot}{IoT}{Internet of Things}
\newacronym{opcua}{OPC UA}{Open Platform Communications Unified Architecture}
\newacronym{can}{CAN}{Controller Area Network}
\newacronym{lin}{LIN}{Local Interconnect Network}
\newacronym{tas}{TAS}{Time-Aware Shaper}
\newacronym{cbs}{CBS}{Credit-Based Shaper}
\newacronym{tsn}{TSN}{Time Sensitive Networking}
\newacronym{taprio}{TAPRIO}{Time Aware Priority Shaper}
\newacronym{etf}{ETF}{Earliest Time First}
\newacronym{mqprio}{MQPRIO}{Multiqueue Priority Qdisc}
\newacronym{cots}{COTS}{Commercial off-the-Shelf}
\newacronym{rtt}{RTT}{Round Trip Time}
\newacronym{e2e}{E2E}{End-to-End}
\newacronym{p2p}{P2P}{Peer-to-Peer}
\newacronym{ptp}{PTP}{Precision Time Protocol}
\newacronym{gptp}{gPTP}{generic Precision Time Protocol}
\newacronym{phc}{PHC}{Physical Hardware Clock}
\newacronym{jbod}{JBOD}{Just a Bunch of Devices}
\newacronym{gm}{GM}{Grandmaster Clock}
\newacronym{pubsub}{PubSub}{Publish–Subscribe}
\newacronym{tc}{tc}{traffic control}
\newacronym[plural=TCLs,firstplural=traffic classes (TCLs)]{tcl}{TCL}{Traffic Class}
\newacronym{qdisc}{qdisc}{queuing discipline}
\newacronym{qos}{QoS}{Quality of Service}
\newacronym{ecdf}{ECDF}{Empirical Cumulative Distribution Function}
\newacronym{be}{BE}{Best Effort}
\newacronym{kpi}{KPI}{Key Performance Indicator}
\newacronym{pcp}{PCP}{Priority Code Point}
\newacronym{candc}{C\&C}{Command \& Control}
\newacronym{skb}{SKB}{Socket Buffer}
\newacronym{ivn}{IVN}{Intra-Vehicular Network}
\newacronym{oc}{OC}{Ordinary Clock}
\newacronym[plural=TRCLs,firstplural=Transparent Clocks (TRCLs)]{tclo}{TRCL}{Transparent Clock}
\newacronym{bc}{BC}{Boundary Clock}
\newacronym{pfifo}{PFIFO}{Packet Limited First In, First Out queue}
\newacronym{sut}{SUT}{System Under Test}
\newacronym{adas}{ADAS}{Advanced Driver Assistance Systems}
\newacronym{phy}{PHY}{Physical Layer}
\newacronym{udp}{UDP}{User Datagram Protocol}
\newacronym{bmca}{BMCA}{Best Master Clock Algorithm}
\newacronym{tcp}{TCP}{Transmission Control Protocol}
\newacronym{os}{OS}{Operating System}
\newacronym{irq}{IRQ}{Interrupt Request}
\newacronym{cpu}{CPU}{Central Processing Unit}
\newacronym{gpu}{GPU}{Graphics Processing Unit}
\newacronym{smp}{SMP}{Symmetrical Multiprocessing}
\newacronym{smt}{SMT}{Simultaneous Multi-Threading}
\newacronym{rt}{RT}{Real-Time}
\newacronym{hw}{HW}{Hardware}
\newacronym{sw}{SW}{Software}
\newacronym{dvfs}{DVFS}{Dynamic Voltage and Frequency Scaling}
\newacronym{vlan}{VLAN}{Virtual LAN}
\newacronym{kc}{KC}{Key Contribution}
\newacronym{pcap}{PCAP}{Packet Capture}
\newacronym{utc}{UTC}{Coordinated Universal Time}
\newacronym{tai}{TAI}{International Atomic Time}
\newacronym{sdn}{SDN}{Software Defined Networking}
\newacronym{sdk}{SDK}{Software Development Kit}
\newacronym{api}{API}{Application Programming Interface}
\newacronym{tdma}{TDMA}{Time Division Multiple Access}
\newacronym{ao}{AO}{Automatic Overclocking}
\newacronym{ovs}{OvS}{Open vSwitch}
\newacronym{sriov}{SR-IOV}{Single Root I/O Virtualization}
\newacronym{vm}{VM}{Virtual Machine}
\newacronym{lxc}{LXC}{Linux Container}
\newacronym{lxd}{LXD}{Linux Container Daemon}
\newacronym{vf}{VF}{Virtual Function}
\newacronym{pci}{PCI}{Peripheral Component Interconnect}
\newacronym{tee}{TEE}{Trusted Execution Environment}
\newacronym{cgroup}{cgroup}{Control Group}
\newacronym{oci}{OCI}{Open Container Initiative}
\newacronym{netem}{netem}{Network Emulator}
\newacronym{zkp}{ZKP}{Zero Knowledge Proof}
\newacronym{zk}{ZK}{Zero Knowledge}
\newacronym{ipfs}{IPFS}{InterPlanetary File System}
\newacronym{cft}{CFT}{Crash Fault Tolerant}
\newacronym{bft}{BFT}{Byzantine Fault Tolerant}
\newacronym{tps}{TPS}{Transaction per Second}
\newacronym{sgx}{SGX}{Software Guard Extension}
\newacronym{tdx}{TDX}{Trusted Domain Extensions}
\newacronym{sev}{SEV}{Secure Encrypted Virtualization}
\newacronym{pos}{PoS}{Proof-of-Stake}
\newacronym{pow}{PoW}{Proof-of-Work}
\newacronym{pbft}{PBFT}{Practical BFT}
\newacronym{ba}{BA}{Byzantine Agreement}
\newacronym{avm}{AVM}{Algorand Virtual Machine}
\newacronym{evm}{EVM}{Ethereum Virtual Machine}
\newacronym{dapp}{DApp}{Distributed Application}
\newacronym{dlt}{DLT}{Distributed Ledger Technology}
\newacronym{mev}{MEV}{Maximal Extractable Value}
\newacronym{pbs}{PBS}{Proposer-Builder Separation}
\newacronym{mempool}{mempool}{memory pool}
\newacronym{dkg}{DKG}{Distributed Key Generation}
\newacronym{mflops}{MFLOPS}{Mega Floating-Point Operations Per Second}
\newacronym{ttd}{TTD}{Terminal Total Difficulty}

\newacronym{engine}{EnGINE}{\textit{\textbf{En}vironment for \textbf{G}eneric \textbf{I}n-vehicular \textbf{N}etworking \textbf{E}xperiments}}
\newacronym{smr}{SMR}{State Machine Replication}
\newacronym{methoda}{METHODA}{\textbf{M}ultilayer \textbf{E}nvironment and \textbf{T}oolchain for \textbf{H}olistic \textbf{N}etw\textbf{O}rk \textbf{D}esign and \textbf{A}nalysis}

\section{Introduction}
\label{sec:introduction}
Since the introduction of Bitcoin in 2008~\cite{nakamoto2009bitcoin}, we have seen increased activity in distributed and decentralized systems research and solutions.
The area of Blockchain-based solutions is of particular interest, with the introduction of Layer-1 protocols such as Algorand~\cite{giladAlgorand2017}, Aptos~\cite{AptosWhi38:online}, Cosmos~\cite{CosmoWhitepaper:online}, or Ethereum~\cite{Buterin2013} offering advances and extensions on consensus and execution layers. 
Furthermore, there is a notable emphasis on advanced cryptographic protocols. 
Some of these protocols rely on threshold cryptography for secure private key protection, while also incorporating privacy features through \glspl{zkp} at the application layer for users~\cite{cryptoeprint:2014/349}. 
Recent developments have also focused on scalability, utilizing \gls{zk} rollups to increase \gls{tps}~\cite{thibault2022ZK}.
Additionally, \glspl{tee} are employed for secure computation~\cite{li2022sok}. 
Other innovations include privacy-preserving networks like Nym~\cite{nymwhite38:online} and distributed storage solutions such as the \gls{ipfs}~\cite{benet2014ipfs}.
Often, these heterogenous systems interact on different application stack layers for improved performance, security, usability or privacy.
For instance, \gls{zk} rollups built on top of underlying Layer-1 solutions, \glspl{tee}, or threshold cryptosystems, should run within the consensus mechanisms of a corresponding blockchain~\cite{cryptoeprint:2023/427} or as an off-chain solution. Privacy-preserving networks, e.g., Nym, offer unlinkability on the networking layer for various deployments.

As modern distributed systems continue to evolve, they become increasingly complex. 
To grasp their practical implications and potential improvements, it's crucial to understand each component in detail.
This requires precise control over the deployment environment. 
Such insights can help the core developers of given systems to identify and document relevant parameters for the best performance. Simultaneously, research can identify optimization approaches on individual layers, do system modeling, and investigate particular extensions and their impact on the system.  

When examining cross-layer approaches, we identified that existing frameworks exhibit shortcomings in terms of modularity, upgradeability, and their ability to comprehensively assess broader distributed protocols, like threshold cryptography.
Some solutions tend to focus on specific aspects or a restricted set of metrics and parameters~\cite{nasrulin2022gromit,gramoli2023Diablo}. Notably, prior evaluations of large-scale systems, particularly those tied to cloud deployments~\cite{gramoli2023Diablo}, or those constrained by limited hardware resources for scalability assessments~\cite{nasrulin2022gromit}, may not accurately reflect real-world system dynamics and could introduce artifacts that affect measurement results.

We propose a methodological approach to assess various solutions on a common platform and observe their interactions and possible implications on scenario-specific \glspl{kpi}.
The methodology considers the deployment strategies, suitable experiment design, and systematizes experiment metrics and parameters. 

We focus on local deployments that limit artifacts, affecting the reproducibility and interpretability. Also we need to ensure scalability and versatility, comparable to cloud deployments.
Also, to handle the complexity of large-scale distributed systems, we have to define fitting experiment methodology applicable to current and future versions of the systems. 
This goes in hand with having a tight and granular control not only over the load generation~\cite{gramoli2023Diablo}, but even more importantly, on the cross-layer setup phase of such systems, as the conditions under which a system is tested can affect the results. 

In the related work analysis, we investigate various frameworks towards facilitation of the identified requirements and suitability for future extensions. Rezabek et al.~\cite{Rezabek2022EngineJNSM, RezBosk21} introduced a framework called \engine. 
Its implementation \cite{enginegit} relies on Ansible~\cite{Ansiblei16:online} and 
provides the foundation for basic infrastructure deployments, with an emphasis on finely-tuned experiment setups. 
Its extensibility allows it to not only support current e.g., blockchain systems, but also be future-proof.
This sets it apart from other single-layered frameworks~\cite{gramoli2023Diablo,nasrulin2022gromit}.
While \gls{engine} does have some limitations, which we address in this paper, we recognized its potential for large-scale deployments. 
We recognize \engine~as a suitable base for extension towards large-scale deployments.

In this paper, we present the \sysname, an extension of \gls{engine}.
We integrate \gls{tee} and lightweight virtualization solutions for scalable systems into the base system. 
Since the focus is on distributed solutions, we decided to integrate three sample services - Algorand, Ethereum 2.0, and the FROST~\cite{KomloFrost2020} threshold signature scheme based on Schnorr's algorithm~\cite{crypto-1989-1727}.
We also conduct a baseline evaluation of \gls{tee}-based deployments and combine it with the FROST~\cite{KomloFrost2020} threshold signature scheme.
Lastly, we validate our approach and the \sysname~capabilities by evaluation in a dedicated \gls{hw} infrastructure.
The results show an initial assessment of the capabilities and a verification of the methodological approach.
The \gls{tee} in a combination with threshold protocol introduces negligible overhead on the \gls{e2e} delay, which was already shown by the baseline evaluations. 
Next, we are able to emulate a particular scenario in the blockchain space focusing on \gls{mev} extraction on Ethereum 2.0. 
This can serve as a base for further modelling of such dynamics on Ethereum 2.0 but can be also applied to other solutions.
We do a detailed parameter study for Algorand with evaluation of latency \gls{tps} based on various \gls{hw} specifications and peer's conditions. 
We further publish the codebase, presented results, and additional documentation as an online repository.

\smallskip
Overall we present the following \glspl{kc}:
\begin{enumerate}[label={\bfseries KC\arabic*}, leftmargin=0.87cm]
    \item Methodological approach to assess various distributed systems performance and their possible integration
    \item Detailed analysis of related work focusing on blockchain frameworks
    \item \gls{methoda} - an enhancement to the \engine~ framework for deploying large-scale distributed systems
    \item Comparison of two recent blockchain systems and one threshold signature scheme
    \item Evaluation of a \gls{tee} relying on recent \gls{vm} based solutions
\end{enumerate}

\section{Problem Analysis}
\label{sec:analysis}
This section offers a comprehensive view of the \sysname~distributed system stack. 
We introduce a methodological approach and requirement description to address the challenges we identified in \Cref{sec:introduction}.
Additionally, we conduct a comparative analysis with related studies, quantifying the research gap that motivates our approach. 
Finally, given the plethora of research on evaluating the performance and scalability of blockchain and agreement systems, we distinctly recognize their limitations our work aims to solve.

\begin{figure}[t!]
\centering
\includegraphics[width=\linewidth]{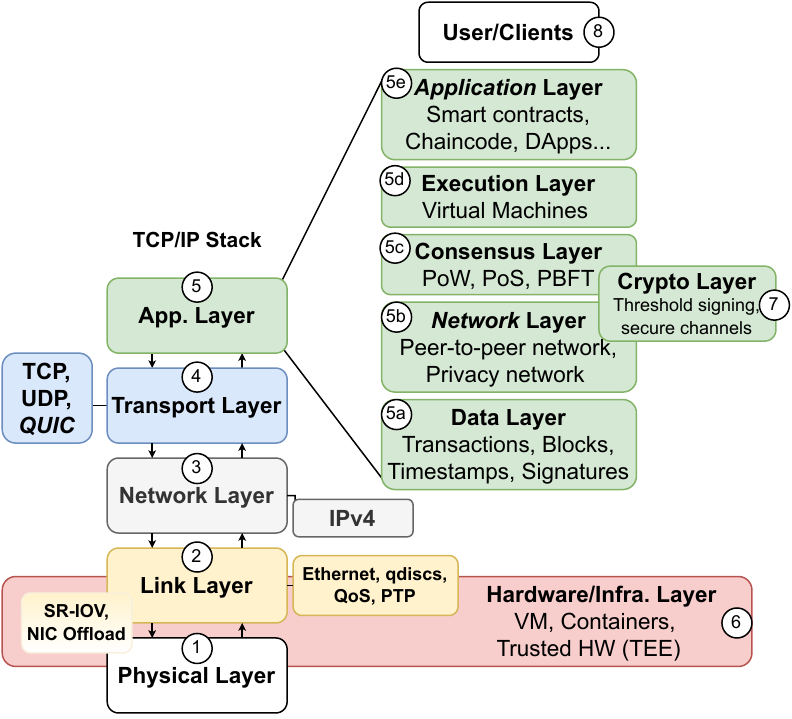}
    \caption{Deployment Stack and Layers, cf. \cite{FanSurvey2020}}
    \label{fig:stackArchitecture}
\end{figure}

\subsection{Distributed Systems Application Stack}
\label{sec:appstack}

Distributed systems, reliant on the TCP/IP stack, facilitate data exchange on a global scale.
As illustrated in \Cref{fig:stackArchitecture}, experimental frameworks must accommodate the entire application and ISO/OSI stack for comprehensive distributed system assessment.
Within this stack, blockchains and the deployment of side-chain or Layer-2 solutions, can be viewed as separate systems. 
These systems need to interact with one another, and our goal is to facilitate this interaction within the proposed framework.
In the following, we provide additional information on layers shown in \Cref{fig:stackArchitecture}.

\encircled{1} refers to the Physical Layer, a supplementary consideration.
Nevertheless, if compatible with upper-layer protocols, the underlying transport medium can be modified. 
In \encircled{2}, Ethernet is envisioned with support for \gls{qdisc} configuration, \gls{ptp}~\cite{linuxptp40} usage on \glspl{nic}, \gls{sriov}, and specific \gls{nic} configuration and offload capabilities. 
\gls{sriov} enables scalability of interfaces and better utilization of the underlying link and \gls{ptp} enables precisely synchronized clocks.
On the Network Layer \encircled{3}, we focus on IPv4 utilization, where IP addresses are directly assigned to corresponding interfaces.
The Transport Layer \encircled{4} should be protocol-agnostic but provide options for execution via e.g., TCP, UDP, or QUIC.

Distributed systems predominantly operate on the Application Layer \encircled{5}, relying on underlying network protocols and infrastructure.
The Data Layer \encircled[0.55mm]{5a} handles various data types, distributed among individual peers in an overlay Network Layer \encircled[0.55mm]{5b} using schemes like gossip or privacy-preserving networks. 
In permissioned and permissionless ledgers, the Consensus Layer \encircled[0.55mm]{5c} deploys Sybil resistance mechanisms (e.g., \gls{pos}, \gls{pow}) and consensus algorithms (e.g., \gls{pbft}~\cite{pbftCastro1999}, Gasper~\cite{buterin2020combining}, or Algorand's \gls{ba}~\cite{giladAlgorand2017}).
Resilience against crash or byzantine failures can be achieved via the \gls{smr} approach \cite{schneider1990implementing}. 
Therefore, for \gls{cft} and \gls{bft} systems, we aim to identify configurations of thresholds under which the system behaves as expected or starts to deteriorate. 
The Execution Layer \encircled[0.55mm]{5d} runs smart contract logic in a corresponding ledger \gls{vm}, e.g., \gls{evm}, \gls{avm}. 
The Application Layer \encircled[0.55mm]{5e} houses the logic of smart contracts and \glspl{dapp}.

To ensure secure data exchange beyond distributed ledgers, incorporating additional distributed protocols like threshold cryptosystems is required. 
Therefore, we introduce a parallel Cryptography Protocol Layer \encircled[0.85mm]{7} positioned between layers \encircled[0.55mm]{5b} and \encircled[0.55mm]{5c}. 
We note that in native deployments (e.g., only running a threshold signature protocol), the additional layers may not carry the same level of importance.

Another essential element is the underlying Hardware and Infrastructure Layer \encircled[0.85mm]{6}. 
Consequently, we not only facilitate bare-metal deployments but also extend support to various lightweight container solutions such as \gls{lxc}~\cite{LinuxCon38:online}, Kata~\cite{KataCont84:online}, or Docker~\cite{DockerAc33:online}. 
While the system may potentially be improved by \gls{lxd}~\cite{Runsyste81:online} for comprehensive \gls{vm} deployments in the future, the current selections offer all requisite features without additional overhead. 
Additionally, the underlying infrastructure provides supplementary features, including the recent integration of confidential computing capabilities via \gls{tee}, as well as other hardware acceleration capabilities provided by the \gls{cpu} or \glspl{gpu}. 
Finally, it is crucial to emulate users/clients in the system \encircled[0.85mm]{8} and ascertain their interaction with it, as they furnish vital insights regarding the load and its pattern.

\subsection{Potential Use-Case Scenarios}

We now introduce some exemplary use cases for \sysname.
For evaluation of both distributed system in general, and Blockchain systems in particular \sysname~facilitates thorough assessment of \gls{p2p} network properties, such as latency, bandwidth, and node connectivity, crucial for a comprehensive understanding of network performance. Such insights can be later used to confirm or model metrics, such as worst-, average-, and best-case latencies, among others. 
In context of \gls{mev} analysis, our solution allows to shed light on transaction ordering strategies, their influence on blockchain security, and the overhead impact of various privacy-preserving techniques.
Furthermore, the framework facilitates examining node operator overhead, offering insights into resource requirements to ensure optimal system scalability and reward calculation. It also allows analysis of communication protocol replacement, extension, and configuration, providing a controlled environment for rigorous experimentation with various networking approaches.

\subsection{Evaluation Methodology}
\label{sec:evaluation_methodology}
This section outlines the methodology we employ to effectively evaluate distributed systems' performance and functionality. 
We discuss deployment strategies, considered experiment design, as well as metrics and parameters.
\subsubsection{Deployment Strategies}
In terms of deployment capabilities, we identify two options -- simulation and emulation.
For emulation we further consider either a private network or data centers (cloud).
Based on the related work, we see that simulators~\cite{singh2008bft,9833775} focus only on particular scenarios and might provide limited insights to cross-layer and infrastructure impact, including networking properties of the system. 
However, their main advantage is a pure focus on a given algorithm's scalability without introducing noise and other artifacts affecting involved peers. 
For the emulation deployments, we see both examples in the related work, e.g., Gromit~\cite{nasrulin2022gromit} relies on private infrastructure, whereas Diablo~\cite{gramoli2023Diablo} on cloud deployments. 
The main argument for cloud deployments is simple access to more computational nodes, which can improve the scalability of the evaluation. 
However, a critical shortcoming is experiment reproducibility, especially for configurations and investigation of layers \encircled{2}-\encircled{4}, as the underlying experiment conditions change depending on too many factors, such as a load in the used data centers, latencies among the networks, and many others.
Such noise and instabilities make collecting precise insights challenging.
In comparison, private networks do not have such issues as all experiments are executed in a controlled environment.
Therefore, our approach focuses on private networks and enabling scalability within them while ensuring tight control over the environment and removing unwanted noise. 
Nevertheless, the framework should be extensible to deployments in the cloud, as it can still be a relevant approach, for instance, long-term probes used to collect data about active systems or single-node experiments.

\subsubsection{Experiment Design}
Our approach uses private network infrastructure to guarantee precise insights and control.
To address scalability, we first consider the experiment design for large-scale systems. 
First, we discuss the applicability of white box and black box testing. 
White box testing requires a detailed code analysis and allows interaction with code to collect new insights, e.g., how expensive a given function is. 
When considering white box testing, introducing code changes is not welcome as in fast-paced environments, e.g., blockchain codebases, introduced changes might soon become obsolete.
As an example of such changes, Ethereum's Geth client~\cite{goethGit} observed 14 releases and more than 800 commits in one year (1 October 2022-2023). 
Similar holds for Algorand's client~\cite{algoGit} with 17 releases.
Nevertheless, white box testing methods are encouraged in the sense of understanding the code's functionality, undocumented behavior, and configuration parameters. 
In case one needs to profile a given code, using external tools, e.g., perf~\cite{PerfWiki38:online}, would be a preferred option. 
On the other hand, black box testing considers only the system's external behavior based on a provided input such as a load. 
Therefore, aiming at black box testing for fast-paced systems without code modification is more suitable. 
As a result, we use white box understanding of the code and use it to improve experiment campaigns while relying on the systems' output to collect relevant insights. 

For a suitable experiment methodology it is essential to address the diversity in hardware capabilities among peers within the system. 
Identifying the minimal hardware specifications that still allow for targeted system performance and sustainable operation is critical. 
Thus, the ability to emulate various configurations becomes crucial for pinpointing the optimal balance between performance and operational costs of the underlying infrastructure. Furthermore, for scenarios where the performance of specific peers in the production system deteriorates, perhaps due to outages or deteriorating network conditions, it is essential to be able to emulate the system's properties under such adverse circumstances. 

To implement these approaches, we consider both local and global perspectives within the system. 
The local view emphasizes individual peer performance, optimizing resource allocation. 
In contrast, the global view assesses overall performance based on collective peer contributions. 
This dual approach offers a comprehensive understanding of system behavior.
Additionally, this strategy necessitates a thorough examination of each layer. 
A feasibility study for every component ensures fair evaluation conditions and provides insights into their combined performance potential.

\subsubsection{Experiment Metrics\,\&\,Parameters}
To comprehensively evaluate a \gls{sut}, we define key metrics and parameters tailored to specific experiments. 
While \gls{tps} and finality are commonly discussed in blockchain contexts~\cite{gramoli2023Diablo,nasrulin2022gromit}, they offer only partial insights. 
To address this, we extend the configuration space to allow for evaluation of additional \glspl{kpi}.
\Cref{tab:metricsParam} outlines eight identified metrics and ten parameters.
Our aim is not to provide an exhaustive list, but to demonstrate possible experiment dimensions when designing an evaluation framework. 
Given the rapid evolution of systems, periodic updates to measurements and parameters will be essential. Therefore, the evaluation framework should be designed with flexibility and extensibility in mind to adapt to present and future requirements.

\begin{table}[t!]
    \centering
     \setlength{\tabcolsep}{2.2pt}
    \begin{tabularx}{\columnwidth}{p{2.1cm} p{6.1cm}}
    \toprule
        \textbf{Term}     & \multicolumn{1}{c}{\textbf{Definition}}    \\
    \midrule
     \multicolumn{2}{c}{\textbf{Metrics}}            \\
  	 \cmidrule{1-2}
  	 \textit{Throughput} & Rate of executed target operations\\
  	 \textit{Latencies} & Latency between e.g., adding a transaction to a block, round-trip-time, processing time, end-to-end latency,... \\

    \textit{Finality} & \gls{dlt} and Blockchain context, Timeout for trust in transaction fixation. \\
    \textit{Queue size} & Amount of data in a queue/mempool \\
    
    \textit{CPU,RAM,I/O} & Local view compute time, RAM usage, I/O \\
  	 \cmidrule{1-2}
     \multicolumn{2}{c}{\textbf{Parameters}}\\
  	 \cmidrule{1-2}
    \textit{Node number} & Number of participants in different roles \\
    \textit{Thresholds} & Threshold values that determine the system's security and availability guarantees (e.g., amount of redundancy in \gls{bft} context) \\
    \textit{Runtime config} & Config options that determine networking, processing, and protocol versions \\
    \textit{Message size / payloads} & Message size and contents of, e.g., votes, transactions, blocks, signatures \\
    \textit{HW specs} & \gls{hw} specification of system nodes \\
    \textit{Network params} & Applicable to private networks (but also cloud and simulations), component specific configuration of delays, packet loss, ... \\
    \textit{NIC config} & NICs configuration and offloading features \\
    \textit{Load} & Workload generation profile, e.g, requests, transactions and smart contracts \\
    \textit{Fees} & \gls{dlt} and Blockchain context, dynamic (transaction) fee configuration \\
    \textit{Faults} & Introduction of faults to assess \gls{sut} capabilities in edge cases \\
    \bottomrule
    \end{tabularx}
    \caption{Identified Metrics and Parameters}
    \label{tab:metricsParam}
\end{table}

\subsection{Requirements Definition}
\label{sec:requirements_definition}
We formalize \sysname~requirements (\textbf{R}) based on the defined application stack in \Cref{fig:stackArchitecture} and the evaluation methodology introduced in \Cref{sec:evaluation_methodology}. 
The \engine~\cite{Rezabek2022EngineJNSM, RezBosk21} authors introduced 13 relevant requirements, but their definitions must be extended to fit our context and needs, as discussed in~\Cref{sec:appstack}.
First, we cover the requirements introduced by \gls{engine}, which we do not amend.  
The general motivation and focus on \reqi, \reqii, and \reqiii~in experiments is highly applicable to our scope. 
To enable those, deployment in private infrastructure, avoiding (e.g., geodistributed) cloud infrastructure is valuable~\cite{Artifact65:online}.
For usability, we rely on \reqiv~(e.g., open source) solutions that are, and use, openly available components. 
Similarly, to handle large-scale deployments, the experiments must run in \reqvi~(without human interaction) once defined.
Second, we list amended and newly defined requirements. 
Insights about the system's behavior in \reqvii~is essential. 
The framework must offer capabilities to emulate crash and byzantine failures in \glspl{sut} to study their robustness and the effect of relevant configuration parameters (e.g., redundancy).
This directly leads to \reqix.
Starting on the node level, we need to have the option to directly allocate given resources, e.g., number of CPU cores, RAM, or \gls{nic} configurations. 
Continuing up the deployment stack (\Cref{fig:stackArchitecture}), each individual \gls{sut} offers distinct configuration parameters. These encompass e.g., the number of peers within the system, communication protocols between them, or transaction fees in blockchain environments.
It is imperative to permit the adjustment of such parameters to study their implications in real-world deployments. 
The objective is to attain a high level of control over as many aspects of the \gls{sut} as possible.

Experiment scalability (\reqxii) stands as an essential requirement.
Our focus extends beyond support and actual measurement of large node count, also including factors like system load (e.g., \gls{tps}), 
wallet numbers, and the overall volume of requests and interactions within the system. 
Our approach complements real-world deployments, owing to the \reqxiv~of \gls{sut} types, log data formats, and number of experiment metrics it supports.
Unlike centralized applications, distributed systems,
typically lack a global perspective on system state and configuration. 
The operator can gather local logs, traffic loads, and general telemetry. 
However, given the distinct data generated by systems like Algorand or Ethereum, this presents challenges. 
To enable meaningful comparisons between different approaches and systems, a \reqxv~ configuration scheme is an essential criterion. This means that, similar configuration and postprocessing options (e.g., according to an abstract template) can be applied across various \glspl{sut}.
The aim is to minimize variability from configuration differences, allowing results to rather reflect the \glspl{sut} intrinsic capabilities.

Fulfillment of the discussed requirements aids with achieving a set of general evaluation goals.
Given \reqvi, \reqvii, \reqix, and \reqxii, a framework can facilitate experiments mirroring real-world environments, accounting for factors like peer ratios in production versus local setups, corresponding workloads, network characteristics, etc. 
A modular framework architecture allows for easy integration of new \glspl{sut}.
It also facilitates maintaining compatibility to dynamically changing \gls{sut} upstream codebases.
All of these characteristics help acquiring insights into bottlenecks, robustness and performance behavior of target \glspl{sut}.

\subsection{Related Work Analysis}
Over the past years, we have seen many evaluation solutions focusing on classical \gls{bft} systems and permissioned or permissionless blockchains.
In our assessment, we have identified distinct categories of frameworks.
Some are tailored to single systems or their specific families~\cite{gupta2016bft,ailijiang2019dissecting,gai2021dissecting, gai2022scaling,Hyperled4:online,GitHubdr43:online}, while others, like Gromit\cite{nasrulin2022gromit} and Diablo\cite{gramoli2023Diablo}, aim to provide a more generalized evaluation approach for specific blockchains.
An assessment, if these systems facilitate the requirements from \Cref{sec:requirements_definition}, is summarized in \Cref{tab:requirementsComparison}. 

\subsubsection{Classical Agreement-Focused Frameworks}
BFT-Bench introduced in \cite{gupta2016bft} emulates various types of faults, collects metrics such as delay and throughput, and considers underlying system CPU or network utilization.
The authors claim implementation of six BFT protocols, including PBFT~\cite{pbftCastro1999} and Zyzzyva~\cite{kotlaZyzzyva2009}.
Since the source code is not available, actual scalability potential and standardization are unknown. Details on post-processing are not provided. 
In a similar direction, Paxi~\cite{ailijiang2019dissecting} implements Paxos \cite{lamport1998part} and its variants~\cite{PaxiGit}, allowing for linearization checks and framework-level fault injections, for e.g., network and node failures. Presented measurements were conducted in a geodistributed cloud deployment.
More recent works introduce and use the Bamboo~\cite{gai2021dissecting, gai2022scaling} evaluation framework, a rework of the Paxi codebase, to prototype and test chained-BFT protocols. The authors implement support for multiple HotStuff~\cite{yin2019hotstuff} variants as well as Streamlet \cite{chan2020streamlet}. The source code is available online\cite{BambooGit}. Bamboo allows for load definition and distribution of additional configuration files among peers.
It supports a simulation and deploy mode, running on several physical nodes or a single machine. The authors state that their measurements were conducted in a cloud deployment, albeit with all machines placed in the same datacenter.
Lastly, a pure simulation approach is introduced by BFTSim~\cite{singh2008bft} building on the NS-2 network simulator~\cite{TheNetwo29:online}. It allows for measurement of various latencies and throughput and various workloads to be provided. No reference to simulator sources is given.
Tool~\cite{9833775, BFTSimulatorGit} is a more recent, open-source simulation framework, implementing a selection of classical agreement algorithms and variants (PBFT, HotStuff) but also protocols, aimed at large-scale blockchain deployments (Algorand \cite{giladAlgorand2017} Consensus). The system allows additional integration of various attacker types.
\subsubsection{Blockchain-Focused Evaluation Frameworks}~\\
BlockBench~\cite{dinh2017blockbench} introduces a benchmarking framework using smart contract cost. It allows for tunable workloads and collection of various metrics, e.g., scalability and fault tolerance insights~\cite{BlockbenchGit}. It works with \gls{pow} Ethereum and Hyperledger Fabric~\cite{Androulaki_2018}. 
For Hyperledger evaluation the authors introduce Hyperledger Caliper~\cite{Hyperled4:online} that recently added support for Ethereum. A purely \gls{evm} focused solution, starting with Ethereum, is Chainhammer~\cite{GitHubdr43:online}. According to the paper, measurements were conducted in private network of commodity machines.
Gromit~\cite{nasrulin2022gromit} aims to be a generic evaluation framework for seven blockchain systems \cite{GromitGit}. While the authors conducted measurements with emulated network delay, automatable fault injection is not implemented in the framework. Experiments were run in a cloud deployment, with machines located in the same datacenter.
Lastly, the Diablo framework~\cite{gramoli2023Diablo} similarly studies latencies and throughput for a range of Blockchain systems. Additionally, general load profiles, smart contracts, and regular transactions are considered. Measurements were conducted in geodistributed cloud deployment. An open-source implementation is available \cite{DiabloGit}, automatable fault-injection is not implemented.

\begin{table}[t]
  \renewcommand{\arraystretch}{0.3}
  \centering
   \setlength{\tabcolsep}{1.4pt}
  \begin{tabularx}{\columnwidth}{l c c c c c c c c c}
    \toprule
     & \cite{gupta2016bft} & \cite{ailijiang2019dissecting} & \cite{gai2021dissecting} & \cite{singh2008bft} & \cite{9833775} & \cite{dinh2017blockbench} & \cite{nasrulin2022gromit} & \cite{gramoli2023Diablo} & Us\\
    \midrule

    Type            & \unknown      & $\dagger$ & $\dagger$ & $\zeta$   & $\zeta$       & $\ddagger$    & $\ddagger$& $\ddagger$& $\ddagger$ \\
    \midrule

    \sreqi          & \yes          & \no       & \yes      & \yes      & \yes          & \yes          & \yes      & \no       & \yes \\
    \sreqii         & \no           & \no       & \maybe    & \no       & \yes          & \yes          & \maybe    & \no       & \yes\\
    \sreqiii        & \no           & \no       & \no       & \no       & \yes          & \yes          & \no       & \no       & \yes\\
    \sreqiv         & \unknown      & \yes      & \yes      & \unknown  & \yes          & \yes          & \yes      & \yes      & \yes\\
    \sreqvi         & \yes          & \maybe    & \yes      & \yes      & \yes          & \yes          & \yes      & \yes      & \yes\\
    \sreqvii        & \yes          & \maybe    & \maybe    &  \yes     & \yes          & \maybe        & \no       & \no       & \yes\\
    \sreqix         & \maybe        & \maybe    & \maybe    & \unknown  & \maybe        & \maybe        & \maybe    & \maybe    & \yes\\
    \sreqxii        & \unknown      & \maybe    & \maybe    & \yes      & \yes          & \maybe        & \yes      & \yes      & \yes\\
    \sreqxiv        & \maybe        & \no       & \no       & \maybe    & \maybe        & \maybe        & \maybe    & \maybe    & \yes\\
    \sreqxv         & \unknown      & \maybe    & \maybe  & \unknown    & \no           & \maybe        & \maybe    & \maybe    & \yes\\
    \bottomrule
  \end{tabularx}
    \caption{Analysis of Related Work Solutions. \yes\,full, \maybe\,partial, \no\,no, \unknown\,unknown satisfaction, $\zeta$ pure simulation, $\dagger$ Implement algo in framework sourcecode, $\ddagger$ Runs third-party code}
\label{tab:requirementsComparison}
\end{table}
\subsubsection{Additional Systems}
Additionally, we explore evaluations of other systems that our framework aims to support. 
A simulator for Nym is introduced in \cite{piotrowska2021Nym}, which measures the system's performance and models its latency based on the number of mix nodes and modeled anonymity set size. 
In the realm of \gls{tee},~\cite{Wang2022} introduces \gls{tee}-based evaluations of various \gls{vm}-based solutions. For low-level assessments of Zero-Knowledge Proofs (\gls{zkp}), zk-Bench~\cite{cryptoeprint:2023/1503} presents a framework to evaluate \gls{zk} circuits and arithmetic.

\subsubsection{Summary}
The analyzed systems from \Cref{tab:requirementsComparison} come in three types: Pure simulators ($\zeta$), frameworks that require implementation of a target algorithm within the native framework source code ($\dagger$), and frameworks that orchestrate and run third-party code ($\ddagger$). While simulators help avoiding setup-related artifacts, insights into complex, real-world deployments is limited. The best approximation of realistic scenarios can be provided by frameworks that orchestrate original codebases in real-world stacks ($\ddagger$).
Systems, relying on (geodistributed) cloud deployments, potentially offer less strict service guarantees, thus allowing for measurement artifacts and variance. Private network deployments with granular control better facilitate \nreqi-\nreqiii.
Frequent partial satisfaction in \nreqix~results from limited configuration capabilities below the application layer. No competitor to \sysname~offers automated \gls{hw} allocation or configuration.
While only true orchestrators ($\ddagger$) effectively handle \gls{sut} codebases, logs, and post-processing, most competitors focus on a very limited set of evaluation metrics (e.g., throughput and latency), affecting \nreqxiv.
In conclusion, our analysis highlights the demand for a true orchestrator ($\ddagger$), facilitating replicability, that also excels in \nreqix, \nreqxiv, and \nreqxv. This includes the ability to evaluate a wider range of metrics, parameters, and experiment methods. Such a framework should not only address classical \gls{bft} or permissioned and permissionless blockchains but also support a broader spectrum of distributed systems to enable research of potential synergies between them.

\section{Design\,\&\,System Architecture}
\label{sec:design}
To satisfy all the outlined requirements defined in \Cref{sec:analysis}, we outline our design decisions and provide more details about the \sysname~architecture.

\subsection{Requirements Satisfaction}
\label{sec:reqSatisfaction}
To ensure \nreqi, \nreqii, and \nreqiii, we start by clearly outlining the capabilities of the employed \gls{hw} and \gls{sw} versions. 
All experiments are conducted in a controlled environment and, free from additional noise.
Our code repository is publicly accessible.
The \gls{sw} artifacts we used are predominantly open-source solutions and \gls{cots} hardware, ensuring both \nreqiv~and helping to establish  \nreqvi.

\subsubsection{Scalability}

For large-scale deployments, an experiment setup with up to tens or hundreds of nodes is crucial (\nreqxii).
Starting with the Link Layer (\encircled{2}, \Cref{fig:stackArchitecture}), we use \gls{sriov}. 
Unlike traditional virtualized environments, where network traffic passes through the hypervisor, \gls{sriov} bypasses this bottleneck by direct communication with the physical \gls{nic}. 
As a result, it significantly reduces CPU overhead and enhances network performance~\cite{Gall2212LatencyLimbo}. 
This is achieved by partitioning the resources of a physical \gls{nic} to multiple \glspl{vf}, where each \gls{vf} acts as an independent \gls{pci} function with its own configuration space and capabilities, e.g., \gls{qdisc} configurations. 
The \glspl{vf} can be assigned to individual virtualization solutions, granting them direct access to the physical \gls{nic}.

To emulate realistic deployments and increase the number of peers, we introduce, among others, lightweight virtualization on the Infrastructure Layer (\encircled{6}, \Cref{fig:stackArchitecture}).
From many variants among application- or system containers, we selected \gls{lxc} system containers.
\gls{lxc} is an \gls{os} system-level lightweight virtualization method that allows multiple isolated Linux systems, known as containers, to run on a single host. 
Unlike an application container, such as e.g., Docker~\cite{DockerAc33:online}, \gls{lxc}'s lifecycle is longer and are managed similar to \glspl{vm}.
On the other hand, unlike traditional \gls{vm}, \gls{lxc} operates at the kernel level, enabling efficient resource utilization~\cite{wiedner2023container}.
Key components for \gls{lxc} are Linux namespaces and \glspl{cgroup} to create isolated application environments.
Namespaces provide process and network isolation, file system views, and more, while \glspl{cgroup} manage resource allocation, including CPU, memory, and disk I/O.
Each \gls{lxc} container is allocated one of the created \gls{vf} interfaces enabled for the corresponding \gls{nic}.
\gls{lxc} supports a variety of Linux distributions, allowing containers to run different \gls{os} distributions and versions. 
Also, \gls{lxc} can be extended by \gls{lxd} for deployments of full \gls{vm} if needed. 
%
Similarly, to integrate and support \glspl{tee} at scale, we use Kata containers~\cite{KataCont84:online} tailored for the use-case of confidential computing. 
Kata is an open-source project and runs containers in its own lightweight \gls{vm}, leveraging hardware virtualization technology that can be fully deployed in a \gls{tee} environment. 
This approach ensures strong isolation, making it more challenging for potential attackers to compromise the guest or host system and isolation among various containers running on the same system. 
Kata Containers' are compatible with orchestration \gls{sw} like Kubernetes. They are also \gls{oci}~\cite{OpenCont21:online} compliant. 
Therefore, the runtime environment supports various container images, facilitating a smooth transition from or to, e.g., Docker containers.

We rely on the \gls{vm}-based solutions, introduced above, for \gls{tee} integration. The process-based solutions, exemplified by Intel \gls{sgx}, enable the creation of secure enclaves with restricted interaction. In contrast, \gls{vm}-based deployments, such as Intel \gls{tdx} or AMD \gls{sev}, enhance virtual machine security through Secure Nested Paging, encrypting and isolating guest \glspl{vm} from the hypervisor, and supporting nested virtualization. The Kata containers then fully run inside the trusted enclave provided by the underlying CPU, which is, in our case, AMD SEV-SNP~\cite{DBLP:journals/corr/abs-2109-10660}. 
Nevertheless, Kata containers allow for other \glspl{tee} technologies such as the Intel \gls{tdx} once present in \gls{cots}.

\begin{figure*}[h!]
\centering
\includegraphics[width=\textwidth]{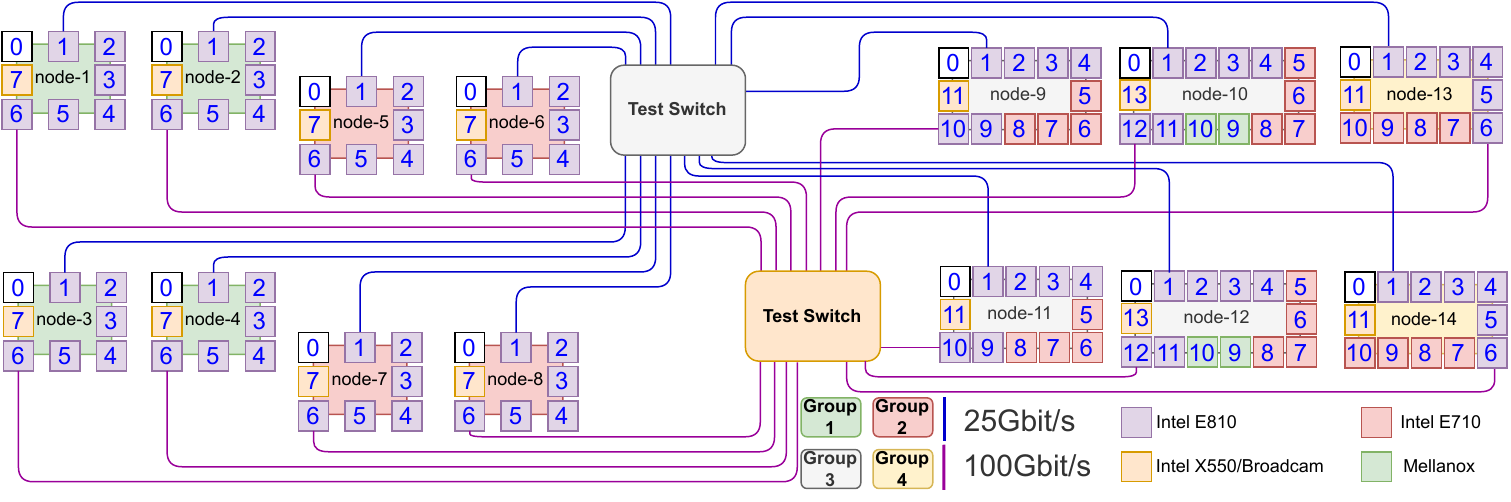}
    \caption{Simplified Topology used in our Deployments, Groups corresponds to \gls{hw} Specification in \Cref{tab:hpclpc}}
    \label{fig:switchTopo}
\end{figure*}

\subsubsection{Realistic Deployments}
Emulating system realism in a controlled environment is a continuous process, especially when the deployed systems evolve. Therefore, it is crucial to have granular control over individual components in the infrastructure. 
In addition, it is important to design experiments that can reflect and properly abstract the complexity and state one can observe in the systems. 
For our experiments, we mainly rely on loads identified by related work and do additional analysis of \emph{mainnet} (Blockchain context: Production network) data for certain  Algorand and Ethereum 2.0 blockchains. 
We deploy several strategies to ensure proper emulation of such setups.

Starting on the Link Layer (\encircled{2}, \Cref{fig:stackArchitecture}), we can emulate various network conditions using the \gls{netem}~\cite{tcnetem817:online} in combination with \gls{mqprio}~\cite{tcmqprio9:online} \gls{qdisc}. The module is controlled by the \gls{tc}~\cite{tc8Linux95:online} functionality of the networking stack. 
\glspl{qdisc} allow prioritizing and manipulating network traffic to ensure shaped transmission. Different \glspl{qdisc} employ various algorithms to determine how they are processed.  
\gls{mqprio} is a \gls{qdisc} module that facilitates multi-queuing with prioritization. It divides network traffic into different queues, each with their own priority level. This allows for fine-grained control over which packets are processed first. Depending on the requirements, it allows for \gls{qos} and combination with additional child \glspl{qdisc}. A relevant child \gls{qdisc} is \gls{netem}, which allows to introduce various network conditions like delay, jitter, packet loss, and packet reordering. By default, the same rule used is applied to all outgoing interfaces. Therefore we combine \gls{netem} with \gls{mqprio}, introducing the capability to modify traffic based on e.g., source IP address or port. We apply various \gls{netem} configurations to each \gls{hw} queue and use nftables~\cite{nftables66:online} to map the corresponding traffic to the relevant \gls{hw} queue on the Application Layer (\encircled{5}). We offer multiple options to configure dedicated resources on each node/peer/container, modify the \gls{nic} configuration, and scale the number of peers. 

Similarly, due to the tight control on the Infrastructure Layer (\encircled{6}, \Cref{fig:stackArchitecture}), we can allocate dedicated resources to each node using either \gls{lxc} or dedicated \gls{nic} configurations to emulate not only delays, but also, for instance, lower the throughput of a given interface and control what \gls{hw} offload capabilities are being used. 
We use native interactions with each system (e.g., provided \glspl{sdk} or native calls on exposed \glspl{api}) and do not rely on any middleware layer. Therefore, any type of interaction can be emulated.

As outlined, a big challenge are versioning and regular changes. 
So, for each protocol we evaluate -- e.g., Algorand and Ethereum 2.0  -- we can also easily select different versions of node and protocol to evaluate and compare their current, previous, and, more importantly, future performance. 
\sysname~supports collection of the metrics defined in \Cref{tab:metricsParam} and due to tightly synchronized clocks, we allow for nanosecond precision on \glspl{nic} that support IEEE 802.1AS~\cite{RezHelm22}. 
We rely on the Linux \gls{ptp} project~\cite{linuxptp40} to handle time synchronization. 
The \gls{ptp} project implements the IEEE 1588 standard for \gls{ptp} and can function over Ethernet, IPv4, or IPv6 networks. The \textit{ptp4l}~\cite{ptp4l8PT0:online} daemon must run on all interfaces, enabling it to synchronize the system's clocks and identify the \gls{ptp} \gls{gm}, which serves as the reference time for the entire system. Additionally, the \textit{phc2sys}~\cite{phc2sys818:online} tool can synchronize clocks within a single node and operate in automatic mode, leveraging information from the \textit{ptp4l} daemon to achieve synchronization.
Due to the precise synchronization, we can inject specific faults and emulate dynamic changes in the system precisely at a given time and observe its effects. 
This allows for valuable insights into the reliability and security of the \gls{sut} (\nreqvii). Furthermore \sysname~supports modification of the various threshold parameters, such as the number of expected active participants, stake distributions in the \gls{pos} systems, waiting parameters, expected synchronicity models in \gls{bft} settings, and others.

As outlined in \Cref{sec:requirements_definition}, satisfying \nreqix~and \nreqxiv~facilitates realistic emulation of complex distributed systems. 
We employ Ansible and YAML for a standardized structure of experiments (\nreqxv) that can be easily ported and to ensure fair comparisons. 
Lastly, to support extentions with new metrics, parameters, and services, we strive for modularity the design of \sysname.
This contributes to easier maintenance and compatibility to upstream changes of \glspl{sut}.

\begin{table*}[tb]
\centering
    \renewcommand{\arraystretch}{0.2}
    \caption{Node families used for experiments with \gls{hw} specs, their count, and (relevant) supported standards by \glspl{nic}}
    \label{tab:hpclpc}        
        \begin{tabular*}{\textwidth}{lccccc}
\toprule
  & \textbf{CPU (cores/threads)} & \textbf{RAM} & \textbf{NICs} \\
    \midrule
Group 1 (4x)   & 24C/48T Intel\textsuperscript{\tiny\textregistered} Xeon Gold 6312U  & 512\,GB DDR4                                                     & 4$\times$\,25 GbE E810-C$^\dagger$, 2$\times$\,100 GbE E810-XXV$^\dagger$ \\
\vspace{2mm}
 & & & 2$\times$\,10 GbE X552$^\dagger$    \\
Group 2 (4x)    & 32C/64T AMD EPYC 7543 & 512\,GB DDR4 & 4$\times$\,25 GbE E810-C$^\dagger$, 2$\times$\,100 GbE E810-XXV$^\dagger$ \\
\vspace{2mm}
& & & 2$\times$\,10 GbE BCM574$^\dagger$  \\
Group 3 (4x)        & 32C/64T AMD EPYC 9354 & 768\,GB DDR5 & 4$\times$\,25 GbE E810-C$^\dagger$, 2$\times$\,100 GbE E810-C$^\dagger$ \\ 
\vspace{2mm}
& & & 2$\times$\,10 GbE BCM574$^\dagger$, 4$\times$\,10 GbE X710$^\dagger$ \\
Group 4 (2x)        & 32C/64T Intel\textsuperscript{\tiny\textregistered} Xeon Gold 6421N & 512\,GB DDR5 & 2$\times$\,100 GbE E810-XXV$^\dagger$, 2$\times$\,100 GbE MT28908$^\dagger$ $^\ddagger$ \\
\vspace{2mm}
& & & 2$\times$\,10 GbE X552$^\dagger$, 4$\times$\,10 GbE X710$^\dagger$, 4$\times$\,25 GbE E810-C$^\dagger$ \\
 
  	 \midrule
\textbf{Total capacity} & \textbf{416C/832T} & \textbf{8192\,GB} & \textbf{On average 9 ports per node}  \\
\bottomrule
\multicolumn{4}{l}{$^\dagger$IEEE 802.1AS, E810 family, E710, X552, and I210 are manufactured by Intel\textsuperscript{\tiny\textregistered}} \\
\multicolumn{4}{l}{$^\ddagger$ Only two out of four nodes have this \gls{nic}}\\                                                     
\end{tabular*}
\end{table*}

\subsection{Integration into \engine}
\begin{figure}[t!]
\centering
\includegraphics[width=\linewidth]{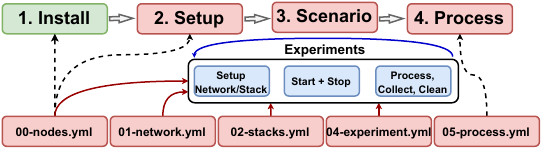}
    \caption{Overview of Extended Modules (highlighted in red) to EnGINE Framework~\cite{BoskRez23IFIP}}
    \label{fig:EnGINEFlow}
\end{figure}

\engine~ is a key component of a broader orchestration approach for consistent, replicable, and verifiable networking experiments~\cite{Rezabek2022EngineJNSM, RezBosk21}.
It offers an adaptable experiment coordination tool, implemented in Ansible, which can be easily integrated with \gls{cots} hardware. 
Experiments conducted in \engine~ follow a structured process orchestrated through a management node, encompassing four key phases~(\Cref{fig:EnGINEFlow}). The \phaseinstall~ phase involves \gls{hw} machine allocation, configuration, \gls{os} deployment, and is something we did not modify. 
The \phasesetup~ phase installs all necessary dependencies.
The actual measurements are executed in the \phasescenario~phase, covering all experiments within a given campaign.
The phase also includes network configuration, as well as the preparation and execution of applications for traffic generation and data collection. 
Each experiment is initiated and concluded, with this cycle repeating until all defined experiments for the scenario are finished. 
The collected results are then processed in the final \phaseprocess~phase.

We now discuss our modifications of \engine~modules, starting with the \phasesetup~ step. We implemented additional features related to \gls{sriov}, container technologies, and other aspects of virtualiziation. We also extended the experiment flow by relevant dependencies as well as network configuration changes.
We no longer rely on \gls{ovs} since we want to generalize the framework with a focus on functionalities from the Network Transport Layer and above.
We already identified suitable technologies that achieve scalability and realism using the Linux network stack (\Cref{sec:reqSatisfaction}). 

Similarly, the \phasescenario~module now features new experiment campaigns for which we defined all relevant applications and services.
For definition of individual experiments, we extended the \texttt{00-nodes.yml} by new options, related to e.g., the use of containers, but also network modes.
After the configuration changes containers are treated similar to full nodes. 
Through this abstraction, no per-service changes are necessary and all deployments and tasks in the pipeline can be reused. 
The same applies to both \gls{lxc} and Kata containers and Docker, respectively. 
\texttt{01-network.yml} now supports additional flags for interface and extended \glspl{qdisc} configuration for the usage of \gls{netem}. 
\texttt{03-stacks.yml} structure itself has not been extensively modified but instead extended by more than \emph{ten} new services  and additional flags for the processing pipeline.
Lastly, we newly introduced \texttt{05-process.yml}, which contains additional metadata relevant to each scenario and its experiment run.
Combining these aspects, we defined \emph{15} experiment scenarios to collect relevant insights and validate our methodology and approach. We enclose a repository with artifacts and source code.

Last, we extended \phaseprocess~  to support large-scale deployments and easier data access and processing via database storage. This includes various applications and various formats such as packet captures, logs, or .csv. These results are correlated and used for visualizations of our metrics (\Cref{tab:metricsParam}).

\subsection{Infrastructure}
\label{sec:infraandexperiments}
Combining all design decisions, we deploy the framework to a private testbed.
A simplified topology setup is shown in \Cref{fig:switchTopo}, with all 14 nodes interconnected via dedicated test switches. 
The nodes are grouped based on their \gls{hw} specifications and \glspl{nic}. 
\Cref{tab:hpclpc} introduces the \gls{nic} types and each group \gls{hw} specifications in addition to the total capacity of the testbed. 
\Cref{fig:switchTopo} uses colors to differentiate \gls{nic} ports. For clarity, we mark each \gls{nic} family with a corresponding color.
Of note, there is additional wiring between the nodes, but for readability reasons it is omitted. 
The additional connections are essential for the \gls{ptp} synchronization. 
Depending on the system, we can scale our experiments to tens or hundreds of nodes represented by the \gls{lxc} containers, based on the total capacity (\Cref{tab:hpclpc}).

For our experiments, we use Ubuntu 22.04 with a 5.15-lowlatency Linux kernel. This image our base \gls{os} for the Algorand, Ethereum 2.0, and FROST Signature scheme experiments. 
For Ethereum 2.0, we use go-ethereum version 1.20~\cite{goethGit} as an execution client and Prysm as a consensus client 4.0.4~\cite{prysmGit}.
For the \gls{pbs} relay we use version 1.11.5-0.2.3~\cite{builderGit}.
Algorand uses the go-algorand release 3.18.1~\cite{algoGit}. 
The FROST library builds on top of the FROST-Dalek implementation~\cite{dalekGit} but was significantly extended by the communication stack and other parts. 
For Kata we also use Ubuntu, but with a version of 20.04 and Linux kernel version 5.19 with additional AMD patches to allow for deployments in the AMD SEV-SNP enclave~\cite{AMDSEVGit}.
Additional details can be found in the enclosed repository.

\section{Evaluation\,\&\,Validation}
This section introduces the experiments outlined in \Cref{sec:infraandexperiments}, serving as a validation to our outlined methods.

\subsection{Feasibility Study}

In this study, our objective is to analyze the effects of implementing the FROST scheme within a \gls{tee} to enhance the security of individual private key shares. \glspl{tee} are being explored for their potential applications in distributed systems, ranging from privacy preservation to security optimization~\cite{li2022sok}. We conduct separate performance assessments of both \glspl{tee} and the FROST threshold signature scheme, as well as their combined impact. This includes overhead comparisons for Kata containers in \gls{tee} vs. native deployments, and white box vs. black box testing for FROST. We maintain consistent configurations, varying only the virtualization technique. \gls{tee}-related experiments run on Group 3 nodes, while FROST without \gls{tee} runs on Group 2 nodes (\Cref{tab:hpclpc}).

\begin{figure}
\centering
\begin{subfigure}[b]{\columnwidth}
    \includegraphics[width=\columnwidth]{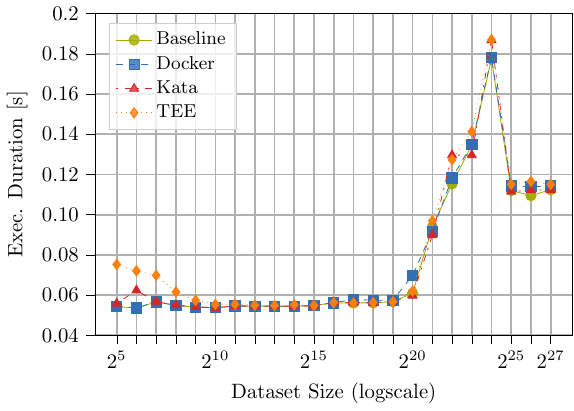}
   \caption{Compute bound CPU Test - Triad Experiment}
   \label{fig:Kata1_ComputeBound} 
\end{subfigure}

\begin{subfigure}[b]{\columnwidth}
    \includegraphics[width=\columnwidth]{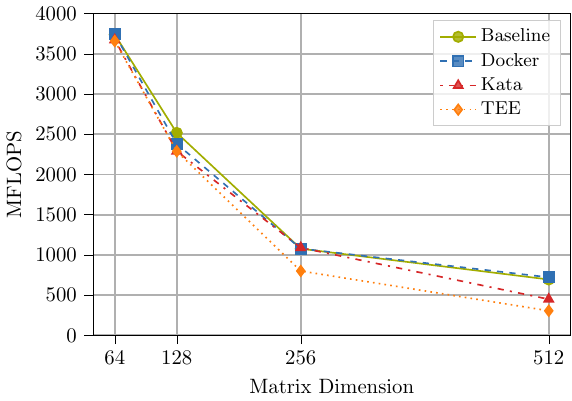}
   \caption{Memory bound CPU Test - Matrix dimension}
   \label{fig:Kata1_Memorybound}
\end{subfigure}
\label{fig:teeEval}
\caption{Comparison of Bare Metal, Docker, Kata, and Kata in TEE - RAM and CPU Tests}
\end{figure}

In initial \gls{tee} experiments, we evaluate the application execution impact. For CPU-bound evaluations, we utilize the \textit{triad} benchmark~\cite{triadWeb}, while matrix multiplication is employed for memory-bound CPU tests, as illustrated in \Cref{fig:teeEval}. The results depicted in \Cref{fig:Kata1_ComputeBound} reveal comparable behavior across baseline (bare metal), Docker and Kata native deployments. Kata within the \gls{tee}, shows a spike for $2^{24}$ elements, attributed to caching effects. Similarly, \Cref{fig:Kata1_Memorybound} displays consistent performance across various matrix dimensions and \gls{mflops} in the memory-bound test, indicating no significant performance degradation for Kata within the \gls{tee}.

\subsubsection{White box and Black box testing}

\begin{figure}
    \centering
    \includegraphics[width=\columnwidth]{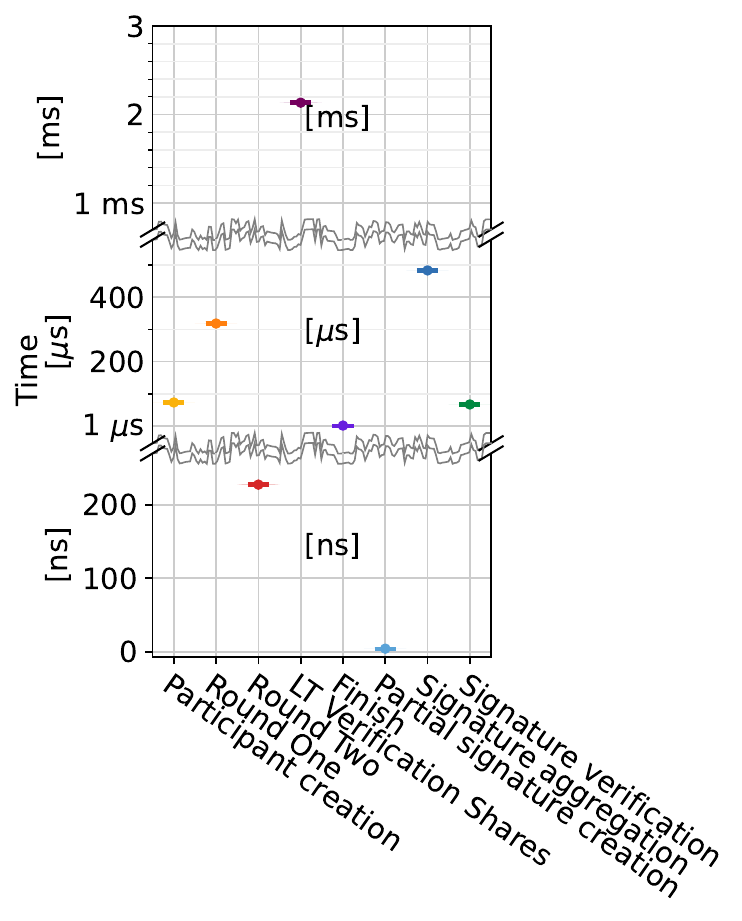}
    \caption{Threshold Schnorr, Whitebox Testing, $n=3$, $t=2$}
    \label{fig:whiteboxSchnorr}
\end{figure}

After establishing the baseline performance of the \gls{tee}, we proceed with the evaluation of Schnorr's signature scheme. This involves employing both white box and black box testing methods to measure the \gls{e2e} latency. The micro-benchmarking is supported by FROST-Dalek~\cite{dalekGit} and results are shown in \Cref{fig:whiteboxSchnorr}. This setup emulates the operations of a distributed scenario on a single node without a networking stack. Notably, nonce generation (preprocessing phase) is excluded. The benchmark breaks down performance into individual steps within each phase, providing a detailed assessment.
Conducted on a Group 2 node (\Cref{tab:hpclpc}), each operation is executed approximately \SI{500}{} times, with $n=3$ and $t=2$.

The \gls{dkg} phase consists of initial operations from \textit{Participant Creation} until the \textit{Finish} step. 
This phase also encompasses the \textit{LT Verification Shares} step, where a participant computes long-term public key shares ($Y_i$) of its peers. Although not strictly part of \gls{dkg}, our system includes it. 
The sign phase involves the \textit{Partial signature creation} that is executed on each peer, followed by \textit{Signature aggregation}—theoretically involving the broadcast of individual partial signatures to all other peers. Finally, the client executes the last step in our system, \textit{Signature verification}. In summary, FROST-Dalek's \gls{dkg} operation, without network communication, takes approximately \SI{2}{\ms}, while threshold signature generation requires around \SI{0.5}{\ms}.

\begin{figure}
\centering
\begin{subfigure}[b]{.75\columnwidth}
    \includegraphics[width=\columnwidth]{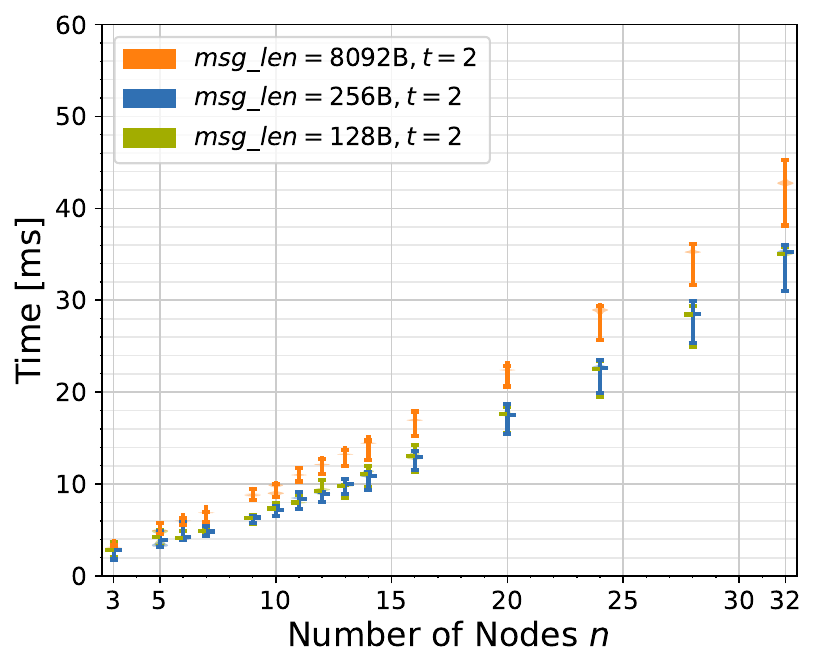}
   \caption{Delay for $n$ and Message Sizes - Without TEE}
   \label{fig:Schnorr_n_notee} 
\end{subfigure}

\begin{subfigure}[b]{.75\columnwidth}
    \includegraphics[width=\columnwidth]{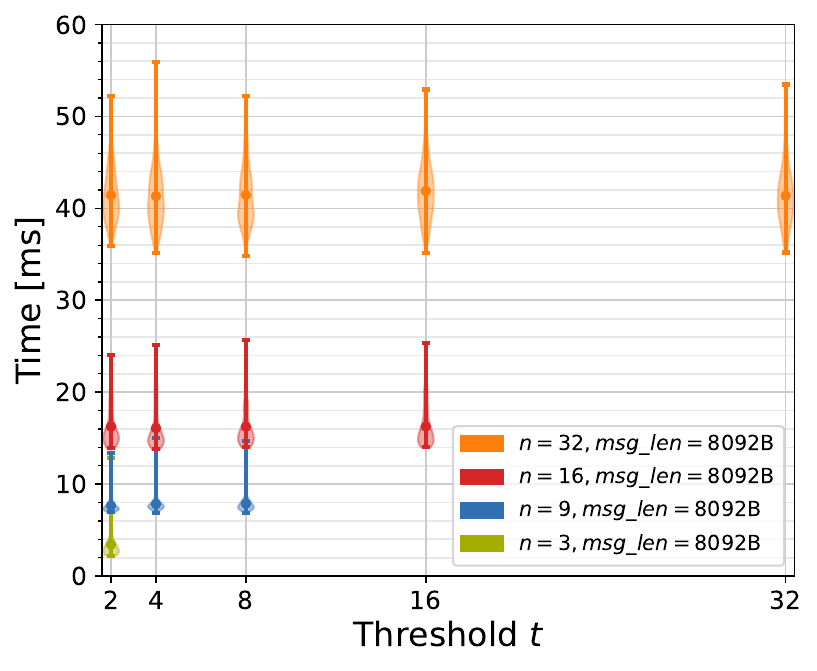}
   \caption{Delay for Various $t$ - With TEE}
   \label{fig:Schnorr_t_tee}
\end{subfigure}
   \label{fig:Schnorr_nt_notee_tee}
\caption{Threshold Schnorr signing End-to-End delay for various Message Sizes, $n$, and $t$ values w/ and w/o \gls{tee}}
\end{figure}

Using black box testing, we assess the \gls{e2e} latencies for signing operation while varying the parameters $n$ and $t$, message size $m$, and execution within or without \gls{tee} (\Cref{fig:Schnorr_nt_notee_tee}). Utilizing hosts from Group 2, we deploy up to eight \gls{lxc} containers per node, each with fixed resources and without threading the signature application. To evaluate the scheme's scalability, we increase the node count to \SI{32}{} and message size up to \SI{8092}{\byte}. Using \sysname s~dynamic fault injection capabilities, we employ \gls{netem} to introduce a delay among peers on the same physical machine, mirroring the delay among the rest of the peers.

In \Cref{fig:Schnorr_n_notee}, we observe that the value of $t$ has a less significant impact on signature generation. This is attributed to its influence being confined to the summation of partial signatures—an operation with low cost (as seen in \Cref{fig:whiteboxSchnorr}). Consequently, the summation of $t$ partial signatures exhibits nearly constant complexity due to the marginal increase in $t$. Similarly, while the differences in means for various message lengths are measurable, they are less critical. Building on the findings from the study of Kata in \gls{tee}, we present \Cref{fig:Schnorr_t_tee}. Here, the mean delay for specific values of $n$, $t$, and message size remains consistent, albeit with more outliers.

In summary, threshold Schnorr, in conjunction with \gls{tee}, achieves mean delays of up to \SI{42}{\ms} for \SI{32}{} nodes and a message size of \SI{8092}{\byte}. Comparing this with \Cref{fig:whiteboxSchnorr}, we observe an approximate 5x increase for signature generation, primarily attributable to communication overhead.

\subsubsection{Applicability to Blockchain}
\begin{figure}
    \centering
    \includegraphics[width=.8\columnwidth]{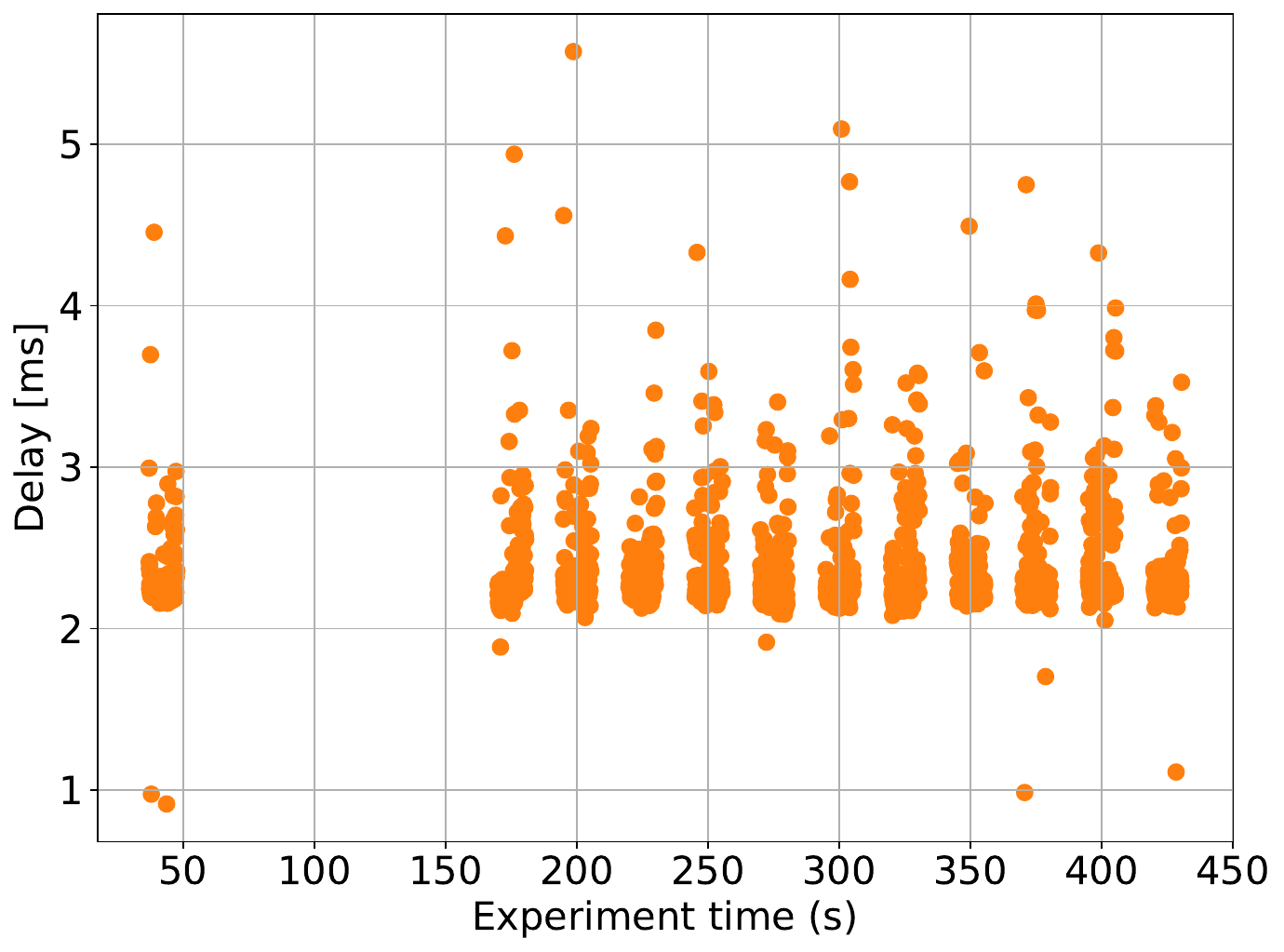}
    \caption{Ethereum 2.0 Client to Mempool \gls{e2e} Latency using tcpdump}
    \label{fig:ethclientmempool}
\end{figure}

From the perspective of a single peer, we are interested in measuring how long a client takes to process a single transaction. Instead of modifying the code of a client node, we want to measure the processing time of a node. Therefore, we measure \gls{e2e} \gls{mempool} confirmation measured on a client node using tcpdump~\cite{mantcpdump}.~\Cref{fig:ethclientmempool} indicates that the majority of transactions exhibit processing delays ranging from \SI{2}{} to \SI{3}{\ms}. Note that no transaction is confirmed between \SI{20}{} and \SI{140}{s}, which is due to the execution clients waiting for the \gls{ttd} to be reached to switch to \gls{pos}~\cite{TTDETh}. A similar approach can be applied to metrics supported by the framework, especially due to the deployment of \gls{ptp} for high precision for raw network data collection.

\subsection{Use-case: Emulation of Ethereum's PBS}
In a specific use-case, we emulate an \gls{mev} scenario of Ethereum 2.0. To mitigate the adverse effects of \gls{mev}, Ethereum 2.0 introduced \gls{pbs}, enabling clients to send transactions directly to block builders~\cite{PBSEth}, instead of validators directly creating blocks. 
Consequently, a double-order auction occurs before a block is dispatched to the proposer, involving three parties: searchers, builders, and relays. Each relay is linked to a group of block builders and selects the most profitable block from the connected set of builders. 
Given that searchers are continually on the lookout for \gls{mev} opportunities, they forward a transaction bundle (a list of transactions) to one or more builders. The builder's goal is to maximize profits by constructing the most lucrative block. This constitutes the first auction, where searchers bid for their bundles in competition with one another. Subsequently, builders compete for the relay's selection of their block, constituting the second auction. We now shift our focus to the first-level auction, specifically on the builder side, to extract the maximum value. We simplify the setup by assuming only a single builder.

\begin{figure}
    \centering
    \includegraphics[width=\columnwidth]{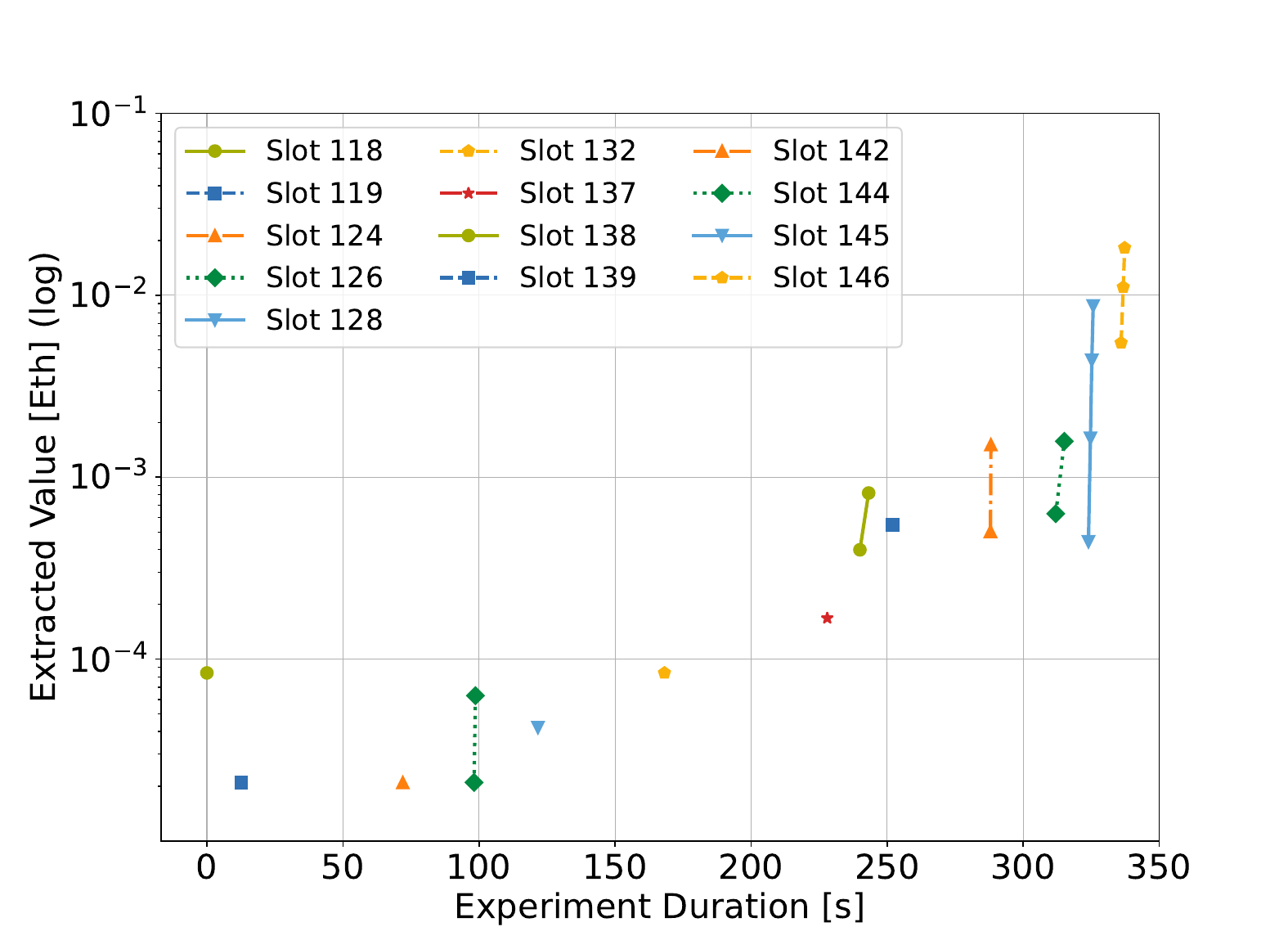}
    \caption{Ethereum Builder-Relay Extracted Value}
    \label{fig:MEV}
\end{figure}

To investigate this behavior, we devise a scenario where a \gls{mev} opportunity is present, leading multiple parties to outbid one another for its extraction. Consequently, certain transactions arriving at the \gls{mempool} have higher gas fees. Out of a load of \SI{20}{\gls{tps}}, \SI{15}{\%} have higher gas fees. The builder then dispatches updated blocks, extracting increased value in each slot, as depicted in~\Cref{fig:MEV}. This value incrementally rises with each new proposed block within the same slot, and cumulatively over the course of the entire experiment. This underscores the builders' objective to generate blocks as frequently as possible while prioritizing transactions with higher gas fees. Overall, this scenario emulates a bidding competition for transaction inclusion in the subsequent block - a basic representation that can be further refined with more sophisticated strategies and the integration of a second-order auction in the future.

\subsection{Optimal Resources}
\begin{figure}
    \centering
    \includegraphics[width=\columnwidth]{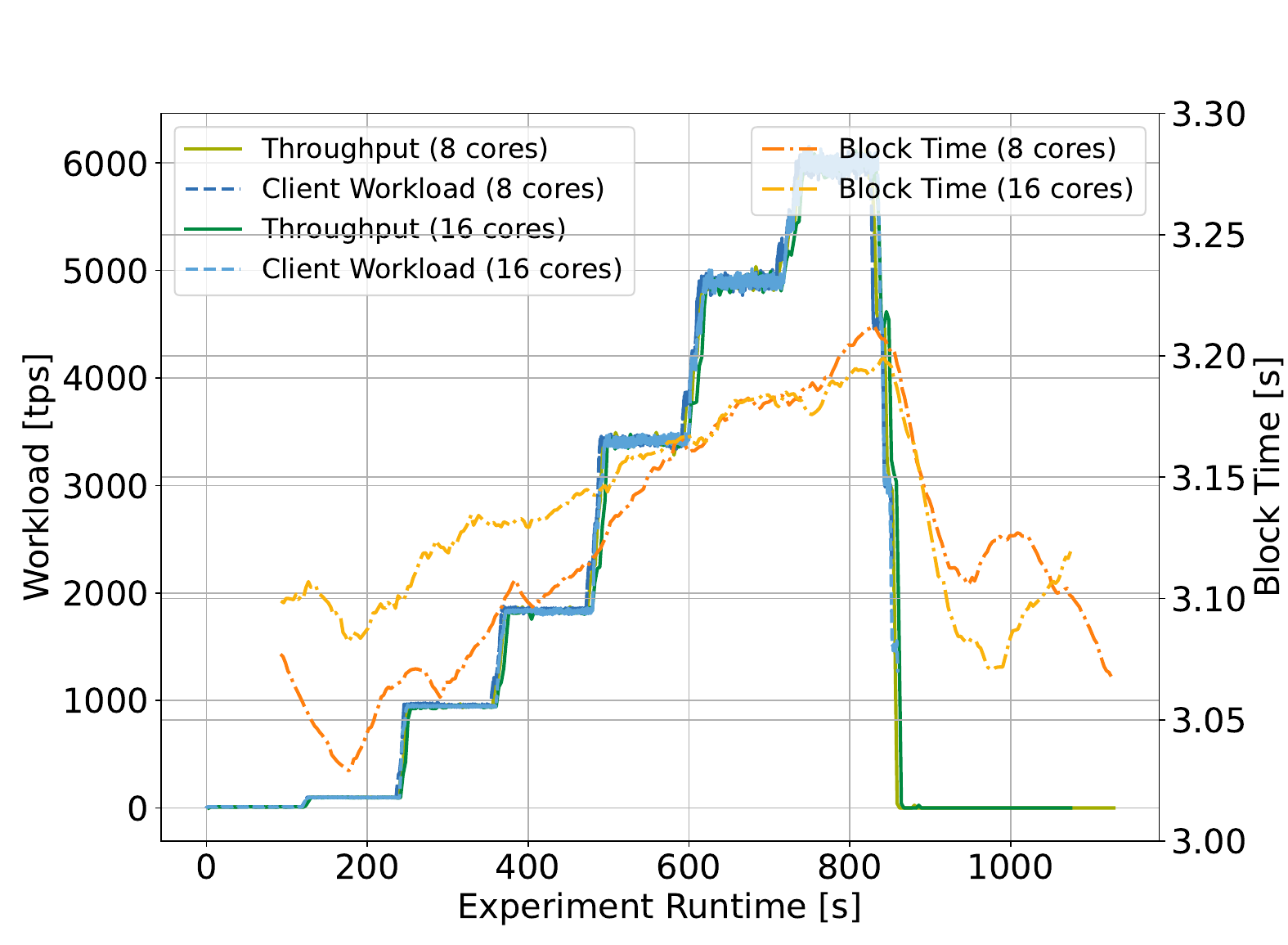}
    \caption{Algorand Throughput under varying Workload and HW Specifications}
    \label{fig:algo-tps}
\end{figure}

Algorand's network differentiates between relay and non-relay (consensus participation and non-participation) nodes, where relays are mainly used to forward traffic to other nodes. Therefore, we investigate the impact of the \gls{hw} specifications of relay nodes on the performance. 
Our Algorand network comprises four relay nodes, eight participation ($\rho$), and eight non-participation ($\eta$) nodes.
Those nodes are distributed as \gls{lxc} containers across four physical nodes with 64 virtual cores each, s.t. each physical node hosts one relay and two \apnode s and \annode s each. Each \annode~has a client container connected to it. 
Those clients generate payment transactions and forward them to the \annode s.
Both \annode s and \apnode s have always eight virtual cores assigned to them. 
For the relays, we consider 8 and 16 virtual cores. 
Algorand recommends eight virtual and 16 cores for $\eta$-/$\rho$- and relay nodes, respectively~\cite{AlgoWeb}. 
\Cref{fig:algo-tps} shows the system's performance under increasing load profiles. 
We observe that both the relay setups can operate loads of up to \SI{6000}{\gls{tps}} without reaching network congestion.
The block time increases slightly with increasing load in both scenarios.
In our restricted setting, we conclude that the number of relay cores does not significantly impact performance. 
In practice, a relay has to handle large numbers of concurrent connections to $\rho$ and $\eta$ nodes, which is not reflected in our experiment.

\section{Challenges}
We encountered several notable challenges while designing \sysname. The testbed's diverse hardware specifications led to varying base performances and capabilities. This diversity is common in real-world deployments, allowing us to conduct assessments that closely resemble realistic scenarios.

For experiment execution, we used a fixed distribution and kernel. Yet, many distributions with potentially distinct performance profiles are common in real-world deployments. Nevertheless, \sysname~facilitates highly-automated and standardized testing of different systems and their respective versions, offering valuable insights into their impact.

\section{Conclusion \& Future Work}
In this study, we unveil \sysname, an extension of the \gls{engine} framework, tailored to heterogeneous, large-scale distributed systems. This advance encompasses a comprehensive requirements and related work analysis, as well as the introduction of a sophisticated application stack and experiment methodology. These components cover a spectrum of experiment strategies and their corresponding metrics and parameters. Additionally, we discuss prerequisites and design choices for \engine~integration in detail.
We conduct a series of experiments to showcase the effectiveness and potential of our methodology within the framework. We present measurements of Ethereum 2.0 and Algorand, the FROST threshold cryptography scheme, and assess performance overhead of processing within a \gls{tee}. \sysname~ offers protocol developers and researchers a holistic approach to acquire in-depth system insights. We validate \sysname~on four \glspl{sut}, collecting \emph{eight} metrics and modifying \emph{12} parameters. For the integration of \gls{tee} with FROST, we observe minimal overhead on the performance with average latency around \SI{40}{\ms}. For Algorand, we see less impact by weaker \gls{hw} specifications on the throughput but rather on latency. Last, emulation of realistic systems behavior e.g., \gls{mev} is possible and could be used to further model such dynamics.  

For future work, we plan to integrate an even broader spectrum of applications, spanning from permissioned and permissionless blockchains to threshold cryptosystems and privacy-preserving networks, among others. Also, while cloud deployments may not be fitting for reproducible experiments, they hold promise as a pertinent infrastructure for sustained data collection, especially with the inclusion of network probes. Consequently, we envision extending the framework to facilitate deployments of such experiments in the cloud.

\bibliographystyle{ACM-Reference-Format}
\bibliography{sample-base}

\end{document}